\documentclass{ieeeaccess}

\usepackage{amsmath,amssymb,amsfonts}
\usepackage[justification=centering]{caption}
\usepackage{algorithmic}
\usepackage{graphicx}
\usepackage{array}
\usepackage{makecell}
\usepackage{textcomp}
\usepackage[colorlinks,linkcolor=blue,anchorcolor=blue,citecolor=blue]{hyperref}
\usepackage[numbers,sort&compress]{natbib}
\def\BibTeX{{\rm B\kern-.05em{\sc i\kern-.025em b}\kern-.08em
    T\kern-.1667em\lower.7ex\hbox{E}\kern-.125emX}}
\begin{document}
\history{Date of publication xxxx 00, 0000, date of current version xxxx 00, 0000.}
\doi{10.1109/ACCESS.2017.DOI}

\title{Quantum Secure Multi-party Summation Based on Entanglement Swapping}
\author{\uppercase{Hong Chang }\authorrefmark{1,2}, \IEEEmembership{Fellow, IEEE},
\uppercase{ Yiting Wu\authorrefmark{1,2}, Gongde Guo\authorrefmark{1,2,3}, and Song Lin,
Jr}.\authorrefmark{1,2,3},
\IEEEmembership{Member, IEEE}}
\address[1]{National Institute of Standards and
Technology, Boulder, CO 80305 USA (e-mail: author@boulder.nist.gov)}
\address[2]{Department of Physics, Colorado State University, Fort Collins,
CO 80523 USA (e-mail: author@lamar.colostate.edu)}
\address[3]{Electrical Engineering Department, University of Colorado, Boulder, CO
80309 USA}
\tfootnote{This paragraph of the first footnote will contain support
information, including sponsor and financial support acknowledgment. For
example, ``This work was supported in part by the U.S. Department of
Commerce under Grant BS123456.''}

\markboth
{Author \headeretal: Preparation of Papers for IEEE TRANSACTIONS and JOURNALS}
{Author \headeretal: Preparation of Papers for IEEE TRANSACTIONS and JOURNALS}

\corresp{Corresponding author: Song Lin (e-mail: lins95@fjnu.edu.cn).}
.

\begin{abstract}
 In this paper, we present a quantum secure multi-party summation protocol, which allows multiple mutually distrustful parties to securely compute the summation of their secret data. In the presented protocol, a semitrusted third party is introduced to help multiple parties to achieve this secure task. Besides, the entanglement swapping of $d$-level cat states and Bell states is employed to securely transmit message between each party and the semitrusted third party. At last, its security against some common attacks is analyzed, which shows that the presented protocol is secure in theory.
\end{abstract}


\titlepgskip=-15pt

\maketitle

\section{Introduction}
\label{sec:introduction}
\par Secure multi-party computing (SMC)\cite{1,2}, which enables $n(n\geq2)$ parties to jointly compute functions based on their private inputs while maintaining the confidentiality of privacy inputs. It is a major branch of modern cryptography, which is widely used in private auctions, secret ballot elections, e-commerce and data mining, and so on. The security of classical SMC is based on the assumption of computational complexity. However, it is increasingly vulnerable with the proposing of quantum algorithms, e.g., Grover search algorithm\cite{3,5}and Shor search algorithm\cite{4,6}. In the meanwhile, Bennett and Brassard proposed a theoretically unconditional secure key distribution protocol (BB84)\cite{7} in 1984, which aroused people's interest in extending classical SMC to the field of quantum mechanics. Recently, various quantum secure multiparty computing protocols have been proposed, such as quantum private query\cite{8,9,10,11}, quantum private comparison\cite{12,13,14,15}, quantum secure multiparty summation\cite{19,20}.

\par Secure multi-party summation is a common secure task in real life, which can be described as follows. There are $n$ participants, and each participant has a secret data. They want to correctly calculate the summation of these inputs without revealing any information about the secret inputs. In 2007, Vaccaro $et$ $al$ first proposed a quantum secure multi-party summation protocol\cite{21}, which was used to implement the secure task of anonymous voting and survey. In this protocol, participants sent secret message to a trusted tallyman, who calculated and published the summation result.
After that, Du $et$ $al$. designed a quantum secret summation protocol \cite{22}based on non-orthogonal single particles, which allows participants to accumulate their private data to an unknown number in a serial manner, so as to achieve the goal of secure multi-party summation. This protocol can resist the collusion attack of $n-1$ participants and achieve asymptotic security.

\par We propose a quantum secure multi-party summation (QSMMS) protocol. Multiple parties that do not trust each other can securely compute the summation of their secret data, while keeping their data private. To achieve this task, these particles of Bell states and cat states are transmitted between participants as signal particles. Eavesdropping detection ensures the security of the transmission of these particles. The secret data of participants are encrypted by entanglement swapping. Finally, a third party is introduced to help all participants obtain the calculation. Here, we require the third party to be semitrusted, that is, she is allowed to conduct herself improperly, but not to collude with either party.

\par The rest of this paper is organized as follows. Some preliminaries are introduced in Sect. 2. We devote Sect. 3 to present the proposed QSMMS
protocol. The security of the proposed protocol is analyzed in Sect. 4. Finally, a short conclusion is provided in Sect. 5.

\section{Preliminaries}
\label{sec:2}
In a $d$-level quantum system, the quantum Fourier transform\cite{23} is a common tool, which can be described as follows,

\begin{equation}
F=\frac{1}{\sqrt{d}}\sum_{j,k=0}^{d-1}\xi^{kj}|j\rangle\langle k|,\label{eq:1}
\end{equation}
where $\xi=e^{\frac{2\pi {\rm i}}{d}}$ and ${\rm i}=\sqrt{-1}$. So, we can construct two common mutually unbiased bases, $B_{1}=\{|j\rangle,j\in D=\{0,1,\cdots,d-1\}\}$ and $B_{2}=\{F|j\rangle, j\in D\}$, directly. Besides  $F$, $X$ and $Z$ are other common operators,

\begin{equation}
X=\frac{1}{\sqrt{d}}\sum_{j=0}^{d-1}|j\oplus 1\rangle\langle j|,\ Z=\frac{1}{\sqrt{d}}\sum_{j=0}^{d-1}\xi^{j}|j\rangle\langle j|,
\label{eq:2}
\end{equation}
where sign $\oplus$ represents the addition module $d$.
The $d$-level Bell state is a generalization form of EPR pairs in $d$-level system (qudits), which  can be expressed as

\begin{equation}
|\Phi(r,w)\rangle=\frac{1}{\sqrt{d}}\sum_{j=0}^{d-1}\xi^{jr}|j\rangle|j\oplus w\rangle,
  \label{eq:3}
\end{equation}
where $r,w\in D$. The state $|\Phi(r,w)\rangle$ can be generated through operations $X$ and $Z$ on particles of $|\Phi(0,0)\rangle=\frac{1}{\sqrt{d}}\sum_{j=0}^{d-1}|j\rangle|j\rangle$,

\begin{equation}
(I\otimes X_{w} Z_{r} )|\Phi(0,0)\rangle=\frac{1}{\sqrt{d}}\sum_{j=0}^{d-1}\xi^{jr}|j\rangle|j\oplus w\rangle=|\Phi(r,w)\rangle,
  \label{eq:4}
\end{equation}
where $Z_{r}=(Z)^{r}$, $X_{w}=(X)^{w}$, and $(X)^{0}=(Z)^{0}=I=|0\rangle\langle0|+|1\rangle\langle1|$. Similarly, the $d$-level cat states can be obtained, which can be described as follows,
\begin{equation}
       |\Psi(u_{1},u_{2},\cdots,u_{n})\rangle=\frac{1}{\sqrt{d}}\sum_{j=0}^{d-1}\xi^{ju_{1}}|j,j\oplus u_{2},j\oplus u_{3},\cdots,j\oplus u_{n}\rangle.
              \label{eq:5}
            \end{equation}
In the proposed protocol, the $d$-level Bell states and the $d$-level cat states are used as information carriers.

\par Entanglement swapping is an important tool in quantum information field. In this paper, we use entanglement swapping between a $n$-particle cat state $|\Psi(v,u,\cdots,u)\rangle$ and $n$ Bell states $|\Phi(r_{i},w_{i})\rangle(i=1,2,\cdots,n)$. When we make a Bell state measurement on the cat state particle and one particle of each Bell state, the remaining particles collapse into an entangled state. Specifically, it can be expressed by the following formula.
\begin{eqnarray}
&&|\Psi(v,u,\cdots,u)\rangle\otimes|\Phi(r_{1},w_{1})\rangle\otimes|\Phi(r_{2},w_{2})\rangle\otimes\cdots\nonumber\\
&&\otimes|\Phi(r_{n},w_{n})\rangle=\frac{1}{\sqrt{d}}\sum_{k,l}\xi^{-kl}|\Psi(\widetilde{v},\widetilde{u}_{1},\widetilde{u}_{2},\cdots,\widetilde{u}_{n})\rangle\nonumber\\
&&\otimes|\Phi(\widetilde{r}_{1},\widetilde{w}_{1})\rangle\otimes\cdots\otimes|\Phi(\widetilde{r}_{n},\widetilde{w}_{n})\rangle,
\label{eq:6}
\end{eqnarray}
where
\begin{equation}
\left\{
\begin{array}{lr}
   \widetilde{v}=v\oplus\sum_{i=1}^{n}k_{i}&\\
   \widetilde{u}_{i}=w_{i}\oplus l_{i}\tag{7}&\\
   \widetilde{r}_{i}=r_{i}\ominus k_{i}&\\
   \widetilde{w}_{i}=u\ominus l_{i}.
 \end{array}
 \right.
\end{equation}
where, sign $\ominus$ represents the subtraction module $d$. Eavesdroppers are unable to extract any useful information from the particles before and after the entanglement. Thus, we can secretly embed private data for encryption operations. In addition, cat state particles and Bell state particles have the following deterministic relationships.\\
\textbf{Theorem 1:}
\par An $(n+1)$-qudit state is in the form of the state $|\Psi(v,u,\cdots,u)\rangle$, when and only when it satisfies both the following two conditions:
\par\textbf{(c1)}when each of its qudits is measured in $B_{1}$ basis, $n + 1$ measurement results satisfy $k_{0}\oplus u=k_{1}=k_{2}=\cdots=k_{n}$;
\par\textbf{(c2)}when each of its qudits is measured in $B_{2}$ basis, $n + 1$ measurement results satisfy $k_{0}\oplus k_{1}\oplus k_{2}\oplus \cdots \oplus k_{n}\oplus v=0$.
Theorem 1 is proved as follows:
\par Simple calculation shows that when $n+1$ qudit is in state $|\Psi\rangle$, each particle is measured with $B_{1}$ or $B_{2}$ basis, and the measurement results meet condition (c1) and (c2).
\par Now, let's prove that if the state $|\Omega\rangle$ satisfies condition (c1) and (c2), then $|\Omega\rangle=\Psi(v,u,\cdots,u)\rangle$.
From condition (c1), we can assume $|\Omega\rangle=\frac{1}{\sqrt{d}}\sum_{j=0}^{d-1}\xi^{jv}|j\rangle|j\oplus u\rangle \cdots|j\oplus u\rangle$, then
\begin{align*}
   &|\Omega'\rangle=X_{u}\otimes I^{n}|\Omega\rangle\\&=\frac{1}{\sqrt{d}}\sum_{j=0}^{d-1}\lambda_{j} X_{u}|j\rangle|j\oplus u\rangle \cdots|j\oplus u\rangle\\&=\frac{1}{\sqrt{d}}\sum_{j=0}^{d-1}\lambda_{j}|j\rangle|j\rangle \cdots|j\rangle\tag{8}
  \end{align*}

  \Figure[t!](topskip=0pt, botskip=0pt, midskip=0pt)[width=4.7 in]{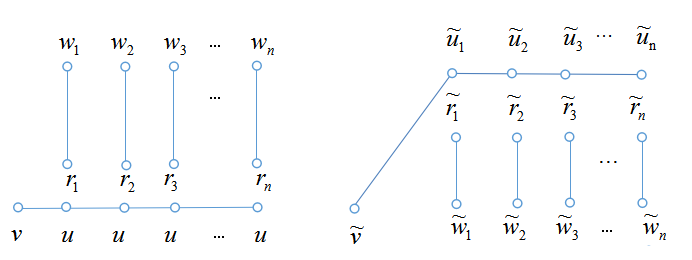}
{The graphical description of entanglement swapping between a $d$-level $n$-particle cat state and $n$ $d$-level Bell states.}

\begin{align*}
&|\Omega'' \rangle= F^{\otimes(n+1)}|\Omega' \rangle\frac{1}{\sqrt{d}}\sum_{j}\lambda_{j}(\frac{1}{\sqrt{d}}\sum_{k_{0}}\xi^{k_{0}j}|k_{0}\rangle)\otimes \cdots\\
&\otimes(\frac{1}{\sqrt{d}}\sum_{k_{n}}\xi^{k_{n}j}|k_{n}\rangle)=\frac{1}{d^{\frac{n+2}{2}}}\sum_{k_{0}\cdots k_{n}}(\sum_{j}\lambda_{j}\xi^{(k_{0}\oplus\cdots+k_{n})j}\\
&|k_{0}\rangle \cdots|k_{n}\rangle.\tag{9}
\end{align*}
where $F=\frac{1}{\sqrt{d}}\sum_{j,k}\xi^{jk}|j\rangle\langle k|,Z_{t}=\frac{1}{\sqrt{d}}\sum_{j=0}^{d-1}\xi^{jt}|j\rangle\langle l|,X_{t}=\frac{1}{\sqrt{d}}\sum_{j=0}^{d-1}\xi^{jt}|j\oplus t\rangle\langle j|$. Because $F X_{t}=Z_{t}F$,
\begin{equation*}
       F^{\otimes(n+1)}X_{u}\otimes I^{n}|\Omega\rangle=Z_{u}\otimes I^{n}F^{\otimes(n+1)}|\Omega\rangle.\tag{10}
\end{equation*}
Since $Z_{u}$ only changes the relative phase and does not change the measurement results of the calculation basis, then the measured results of of the state $|\Omega''\rangle$ in the calculation basis must satisfy condition (c2), which can be obtained $\sum_{j=0}^{d-1} \lambda_{j}\xi^{(k_{0}\oplus k_{1}\oplus \cdots\oplus k_{n})j}=0$ when $k_{0}\oplus \cdots\oplus k_{n}\oplus v\neq0$; $\sum_{j=0}^{d-1}\mid\lambda_{j}\xi^{-vj}\mid^{2}=\frac{1}{d}$ when $k_{0}\oplus\cdots\oplus k_{n}\oplus v=0$.
Thus, we may get $\lambda_{j}=\frac{1}{\sqrt{d}}\xi^{vj}$.
Now we have proved that
\begin{align*}
|\Omega\rangle
       =\frac{1}{\sqrt{d}}\sum_{j=0}^{d-1}\xi^{jv}|j\rangle|j\oplus u\rangle \cdots|j\oplus u\rangle
       |\Psi(v,u,\cdots,u)\rangle.\tag{11}
\end{align*}
\par According to theorem 1, the particle can be guaranteed to be in state $|\Psi(v,u,\cdots,u)\rangle$ through condition (c1) and (c2). Since the Bell state is a special case of cat state, we can get a similar conclusion for the Bell state.\\
\textbf{Corollary 1:}
\par An 2-qudit state is in the form of the state $|\Phi(r,w)\rangle$, when and only when is satisfies both the following two conditions:
\par \textbf{(c1')}when each of its qudits is measured in $B_{1}$ basis, $n+1$ measurement results should satisfy $a_{0}=a_{i}$;
\par \textbf{(c2')}when each of its qudits is measured in $B_{2}$ basis,  the summation of the $n+1$ measurement results should satisfy $a_{0}\oplus a_{i}=0$.

\section{The Proposed Protocol}
\label{sec:3}
In the proposed protocol, there are $n$ mutually distrustful parties labeled $P_{1},P_{2},\cdots,P_{n}$. Each party $P_{i}(i = 1, 2,\cdots,n)$ has a secret dataset $D_{i}=\{x^{1}_{i},x^{2}_{i},\cdots,x^{m}_{i}\}$. We set $S=\{x^{j}_{i}|x^{j}_{i}\in N, 0\leq x^{j}_{i}\leq d-1, i=1,2,\cdots,n, j=1,2,\cdots,m \}$, and then we have $d$ > sup\{S\}. $P_{1},P_{2},\cdots ,P_{n}$ want to jointly compute the summation $\bigoplus_{i=1}^{n} x^{j}_{i}$ with the assistance of a semitrusted third party ( TP), who is allowed to misbehave on her own but cannot conspire with others. Now let us describe steps of the proposed protocol in detail.

\par $\mathbf{(1)}$ Each party $P_{i}(i = 1, 2,\cdots,n)$ prepares $L+\sigma$ copies of $d$-level Bell state.
\begin{equation}
       |\Phi(0,0)\rangle=\frac{1}{\sqrt{d}}\sum_{a=0}^{d-1}|a\rangle_{h^{j}_{i}}|a\rangle_{t^{j}_{i}}.
              \label{eq:8}
\end{equation}
Here, the subscripts $h^{j}_{i}$ and $t^{j}_{i}$ denote two different qudits of a entangled pair, and $j(j=1,2,\cdots,L+\sigma)$ indicates the order of the entangled pairs. $P_{i}$ takes particles $h^{j}_{i}$ and $t^{j}_{i}$ from each pair to form two ordered particle sequences $ H_{i}=(h^{1}_{i},h^{2}_{i},\cdots,h^{L+\sigma}_{i})$ and $T_{i}=(t^{1}_{i},t^{2}_{i},\cdots,t^{L+\sigma}_{i})$. TP prepares $L+n\delta$ copies of $d$-level cat state $|\Psi(v^{j},u^{j},u^{j},\cdots,u^{j})\rangle _{s^{j}_{0}s^{j}_{1}\cdots s^{j}_{n}}$, then takes particle $s^{j}_{i}$ from each cat state $|\Psi(v^{j},u^{j},u^{j},\cdots,u^{j})\rangle$ to construct $n+1$ ordered sequences $S_{0}=(s^{1}_{0},s^{2}_{0},\cdots,s^{L+n\delta}_{0})$, $S_{1}=(s^{1}_{1},s^{2}_{1},\cdots,s^{L+n\delta}_{1})$ ,$ \cdots, S_{n}=(s^{1}_{n},s^{2}_{n},\cdots,s^{L+n\delta}_{n})$, where the superscript $j(j=1,2,\cdots,L+n\delta)$, represents the order of the copies.
\par $\mathbf{(2)}$ TP remains particle sequence $S_{0}$ in her own hands, then sends particle sequence $S_{i}$ to $P_{i}$. Meantime, $P_{i}$ sends particle sequence $T_{i}$ to TP, and remains particle sequence $H_{i}$. The above particles transmission process is shown in Figure 2.

  \Figure[t!](topskip=0pt, botskip=0pt, midskip=0pt)[width=2.5 in]{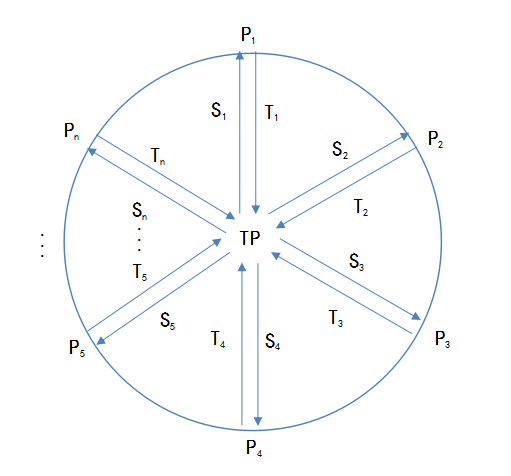}
{The distribution process of TP and $P_{i}$.}
\par $\mathbf{(3)}$ After both TP and participants confirm receiving the particle sequences, the $n+1$ parties (TP and $n$ participants) cooperate to execute the following two eavesdropping detections.\\
$\textbf{Detection 1:}$
\par First, $P_{i}$ randomly chooses $\delta$ qudits from $S_{i}$ as samples. For each sample particle, $P_{i}$ measures it randomly in $B_{1}$ basis or $B_{2}$ basis. Then $P_{i}$ requires TP and the other $n-1$ participants to measure the corresponding qudits in their hands with the same basis. After that, TP announces initial states of sample particles and her measurement result $k_{0}$, then $n-1$ participants announce their measurement results in a random order. According to the announced messages, $P_{i}$ can check whether the measurement results satisfy the conditions (c1) and (c2). In this way, $P_{i}$ can calculate the error rate. Once the error rate is higher than a certain threshold, he will abort the protocol. Otherwise, he will continue.\\
$\textbf{Detection 2:}$
\par First, TP randomly chooses $\sigma$ qudits from $T_{i}$ as samples. For each sample particle, TP measures it randomly in $B_{1}$ basis or $B_{2}$ basis. Then TP requires all $P_{i}$ to measure the corresponding qudits with the same basis. After that, each $P_{i}$ announces his measurement to TP. According to announced messages, TP can check whether the measurement results satisfy the conditions (c1') and (c2').
\par In this way, TP can calculate the error rate, once the error rate is higher than a certain threshold, she will abort the protocol. Otherwise, she will continue.

\par $\mathbf{(4)}$ $P_{i}$ encodes his secret dataset.
Concretely, $P_{i}$ randomly generates two variables $r^{j}_{i} $ and $w^{j}_{i}$ and requires them to meet the condition
$x^{j}_{i}=r^{j}_{i}\oplus\ w^{j}_{i}$. $P_{i}$ performs $Z_{r^{j}_{i}}\otimes X_{w^{j}_{i}} $ operation on the Bell state $|\Phi(0,0)\rangle$. Therefore, the Bell state changes from $|\Phi(0,0)\rangle$ to $|\Phi(r^{j}_{i},w^{j}_{i})\rangle=\frac{1}{\sqrt{d}}\sum_{t=0}^{d-1}\xi^{tr^{j}_{i}}|t\rangle|t\oplus\ w^{j}_{i}\rangle.$

\par $\mathbf{(5)}$ Each participant $P_{i}$ performs the Bell state measurement on particle $s^{j}_{i}$  and particle $h^{j}_{i}$. In this case, the particles $s^{j}_{0},t^{j}_{1},t^{j}_{2},\cdots,t^{j}_{n}$ collapse to a cat state. After that, TP measures this cat state.

\par $\mathbf{(6)}$ After the entanglement swapping, the Bell state of $P_{i}$ becomes $|\Phi(\widetilde{r}^{j}_{i},\widetilde{w}^{j}_{i})\rangle$, where $\widetilde{r}^{j}_{i}=r^{j}_{i}\ominus k^{j}_{i}$
 , $\widetilde{w}^{j}_{i}=u_{i}\ominus l^{j}_{i}$. Then, $P_{i}$ calculates
\begin{equation}
       q^{j}_{i} = \widetilde{r}^{j}_{i}\oplus \widetilde{w}^{j}_{i}= r^{j}_{i}\ominus k^{j}_{i} \oplus u_{i}\ominus l^{j}_{i}
              \label{eq:8}
            \end{equation}
and announces classical information $q^{j}_{i}$ to TP.
\par $\mathbf{(7)}$ After the entanglement swapping, the cat state of TP becomes \\
$|\Psi(\widetilde{v}^{j},\widetilde{u}^{j}_{1},\widetilde{u}^{j}_{2},\cdots,\widetilde{u}^{j}_{n})\rangle$, where $\widetilde{v}^{j}=v^{j}\oplus \sum^{n}_{i=1}k^{j}_{i}$, $\widetilde{u}^{j}_{i}=w^{j}_{i}\oplus l^{j}_{i}$. Then, TP carries out simple calculations and gets

\begin{equation}
       \widetilde{v}^{j}\oplus\sum_{i=1}^{n} \widetilde{u}^{j}_{i}\oplus \sum_{i=1}^{n}q^{j}_{i}\ominus v^{j}\ominus nu^{j}=\sum_{i=1}^{n} r^{j}_{i}\oplus \sum_{i=1}^{n}w^{j}_{i}=\sum_{i=1}^{n} x^{j}_{i}
        \label{eq:9}
\end{equation}
Finally, TP announces the summation to $P_{i}$.

\par Let us consider a simple case as an example to demonstrate the correctness of the presented protocol. Suppose that there are three participants $P_{1}$, $P_{2}$, and $P_{3}$, who have three private datasets $D_{1}=\{1,3\}$, $D_{2}=\{3,6\}$ and $D_{3}=\{2,5\}$, respectively. Thus, we can assume that $d=7>6$. For simplicity, we ignore the eavesdropping checks. Therefore, in step 1, every $P_{i}(i=1,2,3)$ prepares four copies of $d$-level Bell state $|\Phi(0,0)\rangle$, and TP prepares four copies of $d$-level cat state $|\Psi^{1}(4,1,1,1)\rangle,|\Psi^{2}(3,2,2,2)\rangle,|\Psi^{3}(2,5,5,5)\rangle$ and $ |\Psi^{4}(6,3,3,3)\rangle$. Every $P_{i}(i=1,2,3)$ takes particle $h^{j}_{i}$ and $t^{j}_{i}$ from each Bell state to form two ordered particle sequences $H_{i}$ and $T_{i}$. TP takes particle $s^{j}_{i}$ from each cat state to construct four ordered sequences $ S_{0}=(4,3,2,6), S_{1}=(1,2,5,3), S_{2}=(1,2,5,3),S_{3}=(1,2,5,3)$.

\begin{table}
\caption{The proposed protocol in the example.}
\begin{center}
\begin{tabular}{cccccccccc}
  \hline\hline
  $i$ & $j$ & $x_{i}^{j}$ & $r_{i}^{j}$ & $w_{i}^{j}$ & $|\Phi\rangle$ & $k_{i}^{j}$ & $l_{i}^{j}$ & $|\Phi^{'}\rangle$ & $q_{i}^{j}$ \\
  \hline
  1 & 1 & 1 & 1 & 0 & (1,0) & 1 & 1 & (0,1) & 1 \\
  1 & 2 & 3 & 2 & 1 & (2,1) & 1 & 0 & (1,2) & 3 \\
  2 & 1 & 3 & 1 & 2 & (1,2) & 0 & 1 & (1,0) & 1 \\
  2 & 2 & 6 & 4 & 2 & (4,2) & 3 & 1 & (1,1) & 2 \\
  3 & 1 & 2 & 0 & 2 & (0,2) & 0 & 0 & (0,1) & 1 \\
  3 & 2 & 5 & 3 & 2 & (3,2) & 1 & 0 & (2,2) & 4\\
  \hline\hline
\end{tabular}
\end{center}
\end{table}
\par The concrete value of each variable are shown in Table 1, and the new Bell states and $q_{i}^{j}$ can be obtained after the entanglement swapping.
\par In step 7, the specific calculation results are shown in Table 2. It can be seen from this example that the results obtained by TP are correct.

\begin{table}
\caption{The proposed protocol in the example.}
\begin{center}
\begin{tabular}{cccccccc}
  \hline\hline
   j & $|\Psi\rangle$ & $\widetilde{v}^{j}$ & $\widetilde{u}_{1}^{j}$ & $\widetilde{u}_{2}^{j}$ & $\widetilde{u}_{3}^{j}$ & $|\Psi^{'}\rangle$  & $\sum^{3}_{i=1}x_{i}^{j}$ \\
  \hline
   1 & (4,1,1,1) & 5 & 0 & 3 & 2 & (5,0,3,2)& 6 \\
   2 & (3,2,2,2) & 1 & 1 & 3 & 2 & (1,1,3,2) & 0 \\
  \hline\hline
\end{tabular}
\end{center}
\end{table}

\section{Security analysis}
    \label{sec:4}
\par In the following, we analyze the security of the proposed protocol. The security of every quantum channel between TP and participants is ensured by an eavesdropping-check process. Thus, it is evident that stealing information directly is not feasible. An inside attacker is more powerful than an outside attacker. Thus, we focus our attention on the security of protocol against the inside attacks. In addition to dishonest participants, TP is not fully trusted, and she may also steal secret inputs. Thus, two types of attacks will be discussed: attack by dishonest participants and attack by the semitrusted TP.
\begin{center}
  \textbf{A.Attack by dishonest participants}
\end{center}
\par In this attack, we will assume that $P_{m}$ (referred to as $P^{*}_{m}$) is dishonest and wishes to eavesdrop on all or partial information about $P_{l}$'s private data. He can perform one of two common attack strategies to attack the proposed protocol in this article, which are described as follows.
\begin{flushleft}
 \emph{Case 1: Intercept-resend attack }
\end{flushleft}
\par In Step 2, TP transmits sequence $S_{l}$ to $P_{l}$, $P_{l}$ transmits sequence $T_{l}$ to TP. Thus, $P^{*}_{m}$ can utilize this opportunity to execute his attack action. Concretely, $P^{*}_{m}$ prepares several fake particles and replaces the signal particles with these fake particles. First, let's analyze the case where TP transmits sequence $S_{i}$ to $P_{i}$. When sequence $S_{l}$, which contains partial information about $P_{l}$, is transmitted in Step 2, $P^{*}_{m}$ intercepts this particle sequence. After interception, $P^{*}_{m}$ can obtain the value of $u$ in the calculation in Step 6 by measuring the sequence $S_{l}$. However, this action can be detected in Detection 1. Therefore, the attack will fail. Next, let's analyze the other case where $P_{i}$ transmits sequence $T_{i}$ to TP. Similarly, when sequence $T_{l}$ is transmitted in Step 2, $P^{*}_{m}$ intercepts and measures this particle sequence. After then, $P^{*}_{m}$ can obtain the value of $w^{j}_{l}$ in the calculation in Step 6. However, this action can be detected in Detection 2. Therefore, the attack will fail, too. Consequently, this protocol is secure against intercept-resend attack.
\begin{flushleft}
 \emph{Case 2: Entangle-ancilla attack}
\end{flushleft}
\par In this attack, let's first analyze the case where $P_{i}$ transmits sequence $T_{i}$ to TP. Before $P_{l}$ sends sequence $T_{l}$ to TP, $P^{*}_{m}$ prepares an ancilla and performs a certain unitary operation on this ancillary particle and the transmitted particle in the sequence $T_{l}$, which causes the two particles to become entangled. At the end of the protocol, he may utilize this ancilla to eavesdrop on all or some of the information about $P_{l}$. In the following, we will show that even if he has unlimited computing power, with technology limited only by the laws of quantum mechanics, $P^{*}_{m}$ cannot obtain any information about $P_{l}$'s secret input under the condition that no errors are to occur in Detection 2.
\par We can write the most general operation that $P^{*}_{m}$ could apply to $t^{j}_{l}$ particle in sequence $T_{l}$ (particle T) and the ancilla (particle E) as follows:
\begin{align*}
  U:|f\rangle_{T}|0\rangle_{E} \rightarrow \sum^{d-1}_{g=0}|g\rangle_{T}|\varepsilon_{f,g}\rangle_{E}\tag{14}
\end{align*}
where $f\in D$. The states $|\varepsilon_{f,g}\rangle$ are pure ancilla states that are uniquely determined by U. The following conditions can be derived from the unitary nature of U:
\begin{align*}
  \sum^{d-1}_{g=0}(|\varepsilon_{f,g}\rangle, |\varepsilon_{f',g}\rangle)=\delta_{f,f'}\tag{15}
\end{align*}
where $f,f'\in D, \delta_{f,f'}=0 ($or $1)$ when $f\neq f'($or $f=f')$.
\par After $P^{*}_{3}$ performs U operation the quantum system composed of particle H, particle T and particle E will be in the state,
\begin{align*}
 |\Delta\rangle&=U\frac{1}{\sqrt{d}}\sum^{d-1}_{f=0}|f\rangle_{H}|f\rangle_{T}|0\rangle_{E}\\
    &=\frac{1}{\sqrt{d}}\sum^{d-1}_{k=f\oplus g=0}(\sum^{d-1}_{f=0}|f\rangle|f\oplus k\rangle|\varepsilon_{f,f\oplus k}\rangle)_{HTE} .\tag{16}
\end{align*}

\par In Detection 2, we can get from Eq.(L'1)
\begin{align*}
 \sum^{d-1}_{f=0}|f\rangle| f\oplus k\rangle|\varepsilon_{f,f\oplus k}\rangle=\overrightarrow{0},\tag{17}
\end{align*}
where $k=1,2,\cdots,d-1$.
\par According to Eqs.(16) and (17), we can further obtain
\begin{align*}
  |\Delta\rangle&=\frac{1}{\sqrt{d}}\sum^{d-1}_{k=f\oplus g=0}(\sum^{d-1}_{f=0}|f\rangle|f\oplus k\rangle|\varepsilon_{f,f\oplus k}\rangle)\\&=\frac{1}{\sqrt{d}}\sum^{d-1}_{f=0}|f\rangle|f\rangle|\varepsilon_{f,f}\rangle.\tag{18}
\end{align*}
\par When $P_{i}$ and TP measure particles with $B_{2}$ basis, the state of the system can be written as
\begin{align*}
  &F\otimes F|\Delta\rangle=F\otimes \frac{1}{\sqrt{d}}\sum^{d-1}_{f=0}|f\rangle|f\rangle|\varepsilon_{f,f}\rangle\\&
  =\frac{1}{\sqrt{d}}\sum_{k=0}\sum_{y_{H}}\sum_{f}\xi^{fy_{H}\oplus f(k\ominus y_{H})}|y_{H}\rangle|k\ominus y_{H}\rangle)|\varepsilon_{f,f}\rangle.\tag{19}
\end{align*}

\par We get from Eq.(L'2)
 \begin{align*}
   \sum_{y_{H}}\sum_{f}\xi^{fy_{H}\oplus f(k\ominus y_{H})}|y_{H}\rangle|k\ominus y_{H}\rangle)|\varepsilon_{f,f}\rangle=\overrightarrow{0}.\tag{20}
 \end{align*}
where $k=1,2,\cdots,d-1$.
Then,
\begin{align*}
  \sum_{f}\xi^{fk}|\varepsilon_{f,f}\rangle=\overrightarrow{0}.\tag{21}
 \end{align*}

\par After a simple calculation, we can get
\begin{align*}
 |\varepsilon_{0,0}\rangle=|\varepsilon_{1,1}\rangle=\cdots=|\varepsilon_{d-1,d-1}\rangle.\tag{22}
\end{align*}

From Eqs.(20),(21) and (22) we can derive
\begin{align*}
 F\otimes F\frac{1}{\sqrt{d}}\sum^{d-1}_{f=0}|f\rangle|f\rangle|\varepsilon_{f,f}\rangle=\frac{1}{\sqrt{d}}\sum^{d-1}_{f=0}F|f\rangle F|f\rangle\otimes|\varepsilon\rangle,\tag{23}
\end{align*}
where $ |\varepsilon\rangle$ is independent of $|f\rangle$.

\par Above we have deduced $P^{*}_{m}$ successful eavesdropping conditions. When $P^{*}_{m}$ entangles an ancilla particle on a particle T, the probability that $f$ is equal to $g$ is $\frac{1}{d}$. When a sequence has $n$ particles, the probability that $f$ is equal to $g$ is $(\frac{1}{d})^{n}$. As $n$ tends to infinity, the probability of $P^{*}_{m}$ successful eavesdropping tends to 0.
\par For analysis of $P^{*}_{m}$ eavesdropping $S_{l}$, see Appendix 1.

\begin{center}
  \textbf{B.Attack by the semitrusted TP}
\end{center}
\par Now, we will discuss the case in which TP attempts to eavesdrop on private data. In the proposed protocol, TP is allowed to behave improperly, but she cannot conspire with other participants. To get useful information about a participant, TP must get $r$ and $w$ of the participant without introducing any errors into the detection. If TP implements the protocol honestly, then she can obtain $v^{j}$, $u^{j}$, and $w_{i}^{j}$ by measurement in step 2 before entanglement swapping. After entanglement swapping, in step 6, due to the publication of $P_{i}$ , TP can know $q_{i}^{j}$, namely, $\widetilde{r}_{i}^{j}\oplus \widetilde{w}_{i}^{j}$. In step 7, she can obtain $\widetilde{v}^{j}$ and $\widetilde{u}_{i}^{j}$ by measurement after entanglement swapping. At the end of the protocol, she also can know$\sum^{n}_{i=1}x^{j}_{i}$ . Even if TP knew so much information, it could not eavesdrop on the value of $r^{j}_{i}\oplus w^{j}_{i}$. If TP adopts intercept-resend attack and entangle-ancilla attack , the analysis results are the same as above.

\section{Summary and Conclusion}
\label{sec:5}
     \par Before giving a conclusion, we first compare and analyze the proposed with other QMS protocols. In order to highlight the advantages of the proposed protocol more intuitively, we make the following comparisons. Please refer to Table 3 for detailed comparison results. First, several QSMS algorithms were proposed in quantum anonymous voting and surveying protocols in order to calculate the summation of some private data including binary numbers and integers\cite{24}. In 2005, Hillery et al. proposed a QSMS algorithm in their quantum anonymous voting protocol, in which each number is 0 or 1. In 2007, Vaccaro et al. proposed another QSMS algorithm for calculating the summation of $n$ integers, where $n$ denotes the number of parties (voters) in their quantum anonymous surveying protocol, in which a $(n-1)$-particle entangled state is employed. Each party makes a vote by performing a phase-shifting operation on its respective particles\cite{25}. Then all  parties send their particles to the tallyman who is responsible for counting the votes (i.e., calculating the summation of $n$ integers).
\begin{table*}
\caption{ The comparion among proposed protocol and Refs.[24,25]}     
 \centering\begin{tabular}{llll p{5cm}<{\centering}}

  \hline
  \hline
  Protocol&Information carriers&User's operations & Data type \\
  \hline
  proposed protocol&cat states and Bell states&Single operations & Nonnegative integer \\
   Refs.[24]&Cat states and Bell states & Single operations& Binary number\\
   Refs.[25]&N-particle entangled states& Phase-shifting operation & Integer\\
  \hline\hline
\end{tabular}
\end{table*}

 \par From Table 2, we can clearly find the advantages of the proposed protocol. In addition, even if the eavesdropper escaped all detection risks, he would still not have access to useful information.

\par In the proposed  protocol, we propose a QMS protocol that allows multiple parties to securely compute the summation of their secret data. The special feature of the proposed protocol is employing entanglement swapping between cat states and Bell states to encode secret data. The proposed protocols use entanglement swapping between cat states and Bell states to securely transmit message between each party and TP. It is worthy pointing that all parties employ unitary operation to encode their secret data, and generate Bell states. With the help of the semitrusted TP, they can obtain the summation successfully at the end of the protocol.

\section*{Acknowledgment}

This work is partially supported by Fujian Normal University.

\EOD

\end{document}